\documentclass{article}
\usepackage{spconf,amsmath,graphicx}
\usepackage{booktabs} 
\usepackage{graphicx}   
\usepackage{subcaption}
\usepackage{multirow}
\usepackage{tabularx}   
\usepackage{array}
\usepackage{enumitem}

\title{Toward Natural Emotional Text-To-Speech System \\with Fine-Grained Non-Verbal Expression Control}
\name{Wangzixi Zhou, Bagus Tris Atmaja,  Sakriani Sakti}
\address{Nara Institute of Science and Technology, Ikoma, Japan}
\begin{document}
%
\maketitle

\begingroup
\renewcommand\thefootnote{}
\footnotetext{© 2026 IEEE. Personal use of this material is permitted. Permission from IEEE must be obtained for all other uses, in any current or future media, including reprinting/republishing this material for advertising or promotional purposes, creating new collective works, for resale or redistribution to servers or lists, or reuse of any copyrighted component of this work in other works.}
\endgroup

\begin{abstract}

\end{abstract}
%


While current emotional Text-to-Speech (TTS) models have successfully controlled verbal prosody, they often ignore non-verbal vocalizations (NVs), which are essential for authentic human emotion. Although some non-verbal datasets have recently emerged, they often lack high-quality, fine-grained annotations, which restricts a model’s ability to precisely control NV generation. To address this limitation, we propose a novel approach for fine-grained non-verbal expression synthesis. We curate and reprocess female NV utterances from the EARS corpus, develop a new annotation scheme using tags to encode NV types, frequencies, and durations, and build an emotional TTS benchmark to demonstrate its effectiveness. Our evaluation shows that while our NV approach leads to minor trade-offs in perceived naturalness, it significantly improves expressiveness (eMOS 4.20) and emotional recognition accuracy (78.8\%). Emotion-specific analysis further reveals that NV cues are highly effective for high-arousal emotions like happy (82.5\%) and fear (82.7\%), and almost perfectly convey sadness (98.3\%).



\section{Introduction}
\label{sec:intro}

Emotional Text-to-Speech (TTS) synthesis is a growing research area driven by the increasing integration of conversational AI into daily life. Users now expect emotionally rich and engaging responses from dialogue systems, making the ability to synthesize speech with specific emotional tones crucial for creating more empathetic and human-like AI experiences. While existing emotional TTS systems have made significant strides in generating emotionally speech, they have primarily concentrated on manipulating the verbal content. 
These models typically operate by using emotional embeddings to control prosodic features—such as pitch, duration, and energy—derived from the linguistic content. 
While this approach has been successful in many respects, it often ignores a critical component of expressive human communication: non-verbal vocalizations (NVs). As highlighted by Mehrabian~\cite{mehrabian2017nonverbal}, NVs refer to vocal expressions that carry no linguistic content, such as laughter, screams, and sighs. These vocalizations are pervasive in spoken interaction and are known to play a central role in conveying affective states~\cite{scherer1994affect}. They provide a layer of spontaneous, authentic emotional expression that verbal prosody alone cannot capture. Therefore, to achieve truly expressive and human-like emotional TTS, the inclusion and accurate synthesis of NVs are indispensable.

In response to this need, several non-verbal vocalization corpora have emerged. For instance, the AMI Meeting Corpus~\cite{kraaij2005ami} provides 100 hours of meeting recordings with annotations for laughter and coughs. Similarly, a recent work on NVTTS~\cite{borisov2025nonverbaltts} introduced a 17-hour dataset annotated with 10 types of NVs. However, a key limitation of these datasets is their simplified annotation schemes, which typically rely on coarse-grained tags like \textless laugh\textgreater. Such basic annotations severely restrict the model's ability to achieve fine-grained control over NV generation. For instance, a user might wish to control not just the presence of laughter, but its duration and frequency. The lack of a system for precise control over these NV properties remains a significant obstacle to developing advanced emotional TTS models. The ability to specify a cry's duration or the number of discrete laughs within a single utterance is currently unfeasible with existing annotation.

In this paper, we address this limitation by proposing a Fine-Grained Non-Verbal Expression Dataset designed to support more expressive speech synthesis. Our contributions are as follows: 

\vspace{-0.1cm}
\begin{itemize}

\item Fine-grained NV expression data construction: We reprocessed NV utterances from the EARS corpus, focusing on female speakers. We developed a novel annotation scheme for transcription and special tags to enable fine-grained control over NV properties, including style, frequency, and duration.
\vspace{-0.1cm}
\item Fine-grained NV emotional TTS: We built a new NV emotional TTS. The TTS is designed to specifically assess the model's ability to generate speech with fine-grained NV expressions, demonstrating the superiority of our data design in improving both expressiveness and emotional recognition accuracy.

\end{itemize}
\noindent Speech samples are available on the demo page: \footnote {https://37integer.github.io/FINE-GRAINED-NON-VERBAL-TTS/}
\begin{table*}[t]
    \centering
    \caption{The specific transcripts used for each non-verbal category and the total count of each type of utterance within our data.}
    \label{tab:transcript}
    \begin{tabular}{lcccccc}
        \toprule
        \textbf{Category} & \textbf{Cheering} & \textbf{Yelling} & \textbf{Laughter-open} & \textbf{Laughter-closed} & \textbf{Crying} & \textbf{Screaming} \\
        \midrule
        \textbf{Transcript} & 'Wo ho', 'Yo' & 'Hey' & 'Ha' & 'Ha' & 'Whep', 'Wuu', 'Sneeze' & 'Ah' \\
        \textbf{Type Count} & 2 & 1 & 1 & 1 & 3 & 1 \\
        
        \textbf{Count} & 262 & 328 & 266 & 220 & 230 & 154 \\
        \bottomrule
    \end{tabular}
\end{table*}

\section{Related work}
\subsection{Non-verbal Dataset}

While several non-verbal vocalization (NV) datasets have been proposed, their limitations hinder the development of fine-grained, controllable emotional TTS models.

The NVTTS dataset~\cite{borisov2025nonverbaltts}, derived from sources like VoxCeleb~\cite{nagrani2020voxceleb} and Expresso~\cite{nguyen2023expresso}, suffers from poor acoustic quality. With only 1525 of its 3642 utterances being noise-free, it is challenging for smaller TTS models to learn effectively without reproducing acoustic artifacts.
The AMI Meeting Corpus~\cite{kraaij2005ami} is limited by its narrow scope of NVs (mostly laughter and coughs), its domain-specific context of meetings, and its use of predominantly non-native English speakers. Similarly, the Japanese JNV dataset~\cite{xin2024jnv} is unsuitable for English TTS due to its small scale and culturally specific expressions, such as the angry NV "onore."

Therefore, the limited NV scope and cultural specificity of existing datasets necessitate a new, high-fidelity resource for training fine-grained, controllable NV-capable emotional TTS models.

\subsection{Emotional TTS with NV}
Recent advancements in Emotional TTS have seen the integration of NVs, facilitated by Large Language Models (LLMs) and proprietary large-scale datasets. These models allow for a more natural understanding and generation of NVs. For instance, CosyVoice2\cite{du2024cosyvoice} is a streaming TTS system that adds coughs and sighs to its repertoire and offers multilingual synthesis with breathing and laughing capabilities. However, while its architecture is open-sourced, its crucial instruction dataset remains proprietary. Building upon this, NVTTS\cite{borisov2025nonverbaltts} fine-tunes CosyVoice2 on their corpus, enabling the synthesis of a wider variety of non-verbal sounds. This reliance on either proprietary data or fine-tuning large, pre-existing models does not bring substantial technical improvements to conventional TTS systems that lack access to such resources.

\section{Construct Fine-Grained Non-Verbal Expression Data}

Our dataset is constructed by annotating and filtering the existing EARS dataset \cite{richter2024ears}. This section provides an overview of the original data sources and presents statistics on the non-verbal sounds (NVs) and emotion tags in our final dataset.

\subsection{Data Source}
We utilized the EARS dataset, a high-quality speech corpus featuring 107 speakers from diverse backgrounds. The dataset encompasses a wide range of speaking styles, including emotional speech, various reading styles, non-verbal sounds, and conversational free-form speech. For our study, we specifically filtered for non-verbal sounds from 60 female speakers. The selected non-verbal types include six distinct categories: \textbf{laughter-open}, \textbf{laughter-closed}, \textbf{cheering}, \textbf{yelling}, \textbf{crying}, and \textbf{screaming}. The duration of the original audio files was approximately 10–12 seconds.

\subsection{Dataset Processing}

Given that the original audio files contained multiple instances of continuous non-verbal sounds, we first performed preprocessing using the pydub library, a simple yet effective Python tool for audio manipulation. We segmented each audio file based on periods of silence. The segmentation parameters were set as follows: a silence threshold of -40 dBFS, a minimum silence duration of 200 ms, and a silence buffer of 100 ms to be kept around the segmented clips. This process allowed us to extract 739 individual utterances, each lasting approximately 2–6 seconds, from the initial 360 audio recordings.

\begin{table}[h]
    \centering
    \caption{Comparison of the differences between our proposed fine-grained approach and existing coarse-grained methods.}
    \label{tab:comparison}
    \begin{tabular}{cc}
        \toprule
        \textbf{Coarse-grained} & \textbf{Fine-grained} \\
        \midrule
        \textless crying\textgreater & \textless (crying) wuuuuu whep \textgreater \\
        \midrule
          Simple style-level tag &   Style-level tag \\
         - &   Vocalization type and control \\
         - &   Frequency and duration control \\
        \bottomrule
    \end{tabular}
\end{table}

Subsequently, we used the Whisper \cite{radford2023robust} speech recognition model for initial transcription and then performed manual verification to obtain the final non-verbal transcripts. Our transcription strategy, as detailed in Table \ref{tab:comparison}, differs significantly from traditional methods, which often use a single, coarse-grained tag like \texttt{\textless laugh\textgreater ~}to represent non-verbal sounds. We designed a more granular transcription system to achieve precise control and representation of different vocalizations. Our standardization process included the following principles:

\noindent\textbf{Discrete Vocalizations}: For discrete sounds like 'ha' or `whep', we controlled the frequency by repeating specific syllables. For example, we used \texttt{\textless(Laughter-open) ha ha ha\textgreater ~}or \texttt{\textless(Laughter-closed) ha ha ha\textgreater ~ }to differentiate between types of laughter and control the number of laughs within an utterance.

\noindent\textbf{Continuous Vocalizations}: For continuous sounds such as `Wo ho', `Yo', `Hey', `wuu', or `ah', we controlled the duration by repeating the last character. For instance, a `wuu' for crying might last about 1 second, with each additional 'u' adding approximately 0.2 seconds to the duration, allowing for precise control over the length of the cry.

This fine-grained transcription method enabled us to better understand and leverage the underlying structure of non-verbal sounds. The specific transcription types are detailed in Table \ref{tab:transcript}.

\section{Non-verbal Emotional tts }
We selected Grad-TTS \cite{popov2021grad} as our backbone model, which is recognized for its high-quality synthesis of reading-style speech. To enable emotional synthesis, we enhanced the model with an emotion encoder, allowing it to incorporate emotional embeddings. Following Russell’s circumplex model \cite{russell1980circumplex} of affect, we utilized arousal and valence as continuous emotional labels, which provides the model with more fine-grained control over emotional expression.

\begin{figure}[h]
  \centering
  \centerline{\includegraphics[scale=0.45]{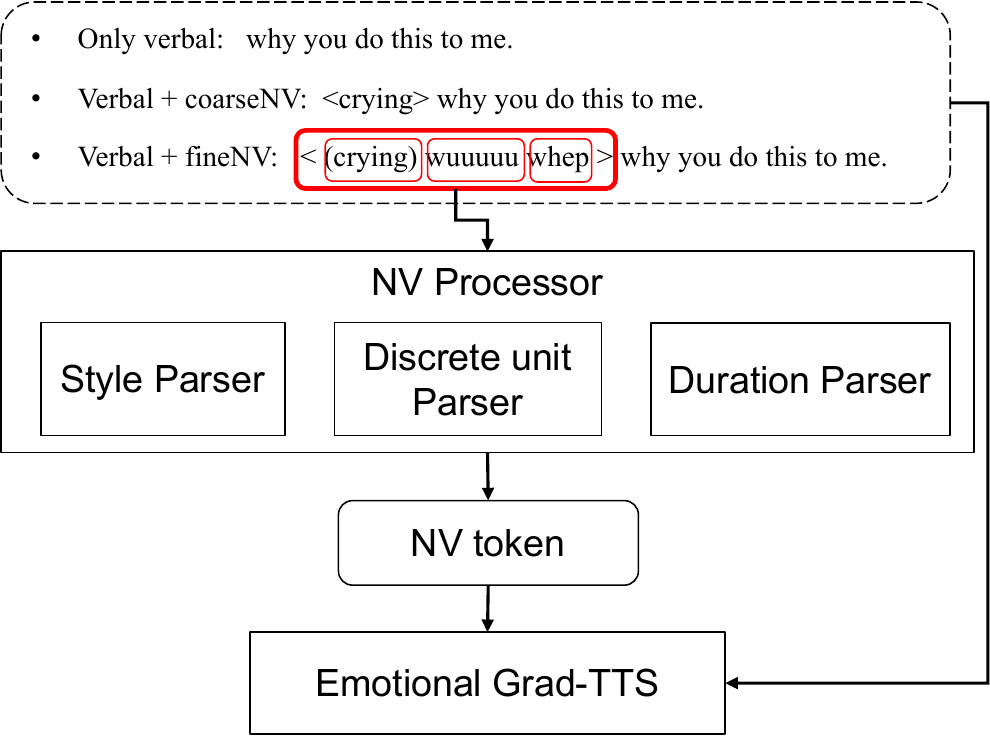}}
\caption{The pipeline of specialized non-verbal processing.}
  \label{fig:model}
\end{figure}

To ensure the model could effectively learn and utilize our specific non-verbal annotations, we customized its text processing pipeline. As shown in the figure \ref{fig:model}, our pipeline is composed of an NV processor that utilizes a style parser, a discrete unit parser, and a duration parser to handle our fine-grained non-verbal expressions. For example, a transcription like \texttt{\textless{(crying) wuuuuu whep\textgreater} why you do this to me} is first processed to identify the non-verbal segment \texttt{\textless{(crying) wuuuuu whep\textgreater}}. This segment is then further analyzed internally: the style Parser identifies the overall style \texttt{crying}, the discrete unit parser counts the occurrences of discrete vocalizations \texttt{whep}, and the duration parser calculates the duration based on continuous vocalizations \texttt{wuuuuu}. These components collectively contribute to the creation of NV tokens that structurally encode these non-verbal sounds, allowing them to be distinctly recognized and processed by the emotional Grad-TTS. This hierarchical and structured annotation method was integrated as the initial step in the text cleaning pipeline, ensuring that the non-verbal information is precisely parsed before being fed into the model. This approach allows the model to synthesize not only verbal speech but also highly expressive, emotionally congruent non-verbal sounds.

\vspace{-0.3cm}
\section{Experienment}
\label{sec:majhead}

\subsection{Training setting}
\vspace{-0.2cm}

To enable the synthesis of emotional verbal speech, we constructed a comprehensive 9-hour mixed dataset of English female speakers, sampled at 22.05 kHz. It was compiled from: EXPRESSO \cite{nguyen2023expresso}, SEMAINE \cite{mckeown2010semaine}, and ESD\cite{zhou2022emotional} datasets. 
To address missing continuous arousal and valence labels in EXPRESSO and ESD, we used a pre-trained Speech Emotion Recognition (SER) model \cite{wagner2023dawn} to predict these values. 
For the non-verbal component, we utilized two distinct datasets to train our model and compare their synthesis performance. We used our curated fine-grained non-verbal expression data to enable precise control over non-verbal vocalizations. For comparison, we also trained the model using the NVTTS corpus \cite{borisov2025nonverbaltts} to represent coarse-grained non-verbal annotations. This allowed for a direct evaluation of our fine-grained approach against a conventional method.

For the acoustic features, we extracted 80-dimensional mel-spectrograms from our processed dataset. These spectrograms serve as the primary input for our model, capturing the necessary frequency and energy information of the speech. We then employed Hifi-GAN \cite{kong2020hifi} as the vocoder, which is responsible for converting these mel-spectrograms back into high-fidelity audio waveforms. The model was trained for 400k iterations on a single GPU (NVIDIA RTX A6000 with 48GB of memory).

\subsection{Evaluation}

\subsubsection{Perceptual and Recognition Evaluation}



We conducted a subjective evaluation with 15 participants to compare three design: (1) verbal-only, (2) verbal + coarse-grained NV (from the ``NVTTS corpus`` which only contained `happy' and `sad' emotions due to a lack of fear/anger-related non-verbal cues) (3) verbal + our proposed fine-grained NV.
To ensure emotion was conveyed through vocal cues rather than text, we chose 20 verbally ambiguous sentences as text input(e.g., "what did you do"). Participants rated 60 samples in total for naturalness (nMOS) and emotion expressiveness (eMOS) on a five-point scale. They also performed a four-choice emotion recognition accuracy task (happy, sad, fear, and anger). Due to data limitations, the ''coarse-grained NV'' was only evaluated for 'happy' and 'sad' NVs.

\subsubsection{Non-verbal Expression Preference Evaluation}

To further refine our non-verbal design, we conducted a Preference Test specifically for happy- and sad-related expressions, as these emotions were associated with multiple non-verbal cues in our dataset.

For happy emotions, we presented participants with four distinct non-verbal expressions combined with an identical verbal utterance:
\vspace{-1mm}
\begin{itemize}[itemsep=0pt,parsep=0pt]
    \footnotesize \item \texttt{<(cheering) Wo ho>}
    \footnotesize \item \texttt{<(cheering) Yo>}
    \footnotesize \item \texttt{<(Laughter-open) ha ha>}
    \footnotesize \item \texttt{<(Laughter-closed) ha ha>}
\end{itemize}
\vspace{-1mm}

Similarly, for sad emotions, we provided four variations combined with the same verbal utterance:
\vspace{-2mm}
\begin{itemize}[itemsep=0pt,parsep=0pt]
    \footnotesize \item \texttt{<(crying) whep>}
    \footnotesize \item \texttt{<(crying) sneeze>}
    \footnotesize \item \texttt{<(crying) wuuuuuuu whep>}
    \footnotesize \item \texttt{<(crying) wuuuuuuu>}
\end{itemize}
\vspace{-2mm}
Participants were asked to rank these combinations based on their preference. 

\subsection{Result and discussion}
\begin{figure}[h]
  \centering
  \centerline{\includegraphics[width=\linewidth]{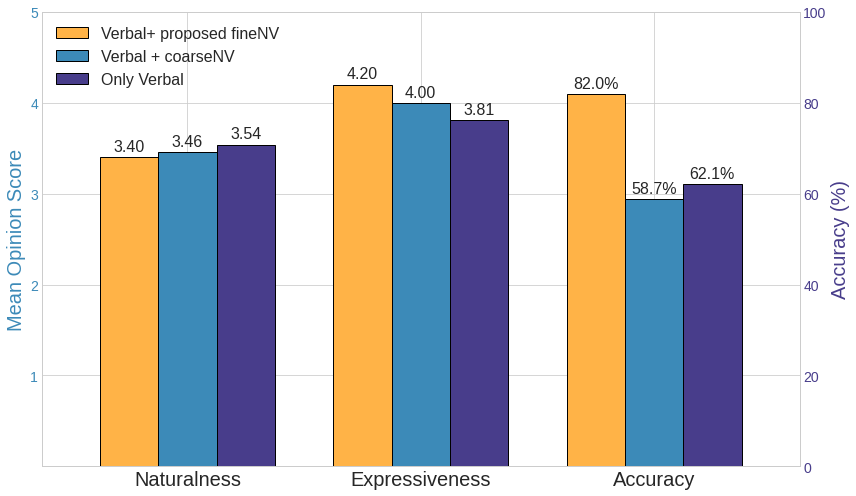}}
\caption{The figure provides performance results of three designs across three key metrics: naturalness MOS, expressiveness MOS, and emotion recognition accuracy.}
  \label{fig:mos}
\end{figure}

The subjective evaluation results are summarized in Figure \ref{fig:mos}. For Naturalness MOS, the "Only Verbal" achieved the highest score of 3.54, suggesting a minor trade-off in perceived naturalness when non-verbal cues are included. In contrast, for Expressiveness MOS, our "Fine-Grained Non-Verbal" significantly outperformed the others with a score of 4.20, demonstrating its effectiveness in enhancing emotional expressiveness.

For emotional recognition accuracy, our method achieved the highest score of 78.8\%, 13.3\% improvement over the ``Only Verbal" baseline. The ``Coarse-grained NV" performed poorly, with low accuracy (especially for 'happy' emotions), indicating its non-verbal cues may have been ambiguous and hindered recognition. A more detailed analysis of this is provided in Section \ref{sec:specific}. These results highlight the primary advantage of our approach: our structured and precise non-verbal cues not only increase perceived expressiveness but also significantly improve the listener's ability to correctly identify emotion.

\begin{table*}[h]
    \centering
    \caption{Detailed nMOS(naturalness mean opinion score) and eMOS(emotion expressiveness mean opinion score)  results for each emotion across the (1) Only Verbal, (2) Verbal + Coarse-grained NV, (3) Verbal + Fine-Grained NV.}
    \label{tab:detailed_results_final_vline}

    \newcolumntype{C}{>{\centering\arraybackslash}X}

    \begin{tabularx}{\textwidth}{ 
        l |         
        *{4}{C} |   
        *{2}{C} |   
        *{4}{C}     
    }
        \toprule
        & 
        \multicolumn{4}{c|}{Only Verbal} & 
        \multicolumn{2}{c|}{Coarse-grained NV} &
        \multicolumn{4}{c}{Fine-Grained NV} \\
        
        \midrule
        
         & Happy & Sad & Anger & Fear & Happy & Sad & Happy & Sad & Anger & Fear  \\
        \midrule
        
        nMOS & 3.67 & 3.69 & 3.61 & 3.19 & 3.19 & 3.73 & 3.43 & 3.67 & 3.34 & 3.18  \\
        eMOS & 3.78 & 3.83 & 3.74 & 3.89 & 3.85 & 4.15 & 4.21 & 4.25 & 4.06 & 4.28  \\
        
        \bottomrule
    \end{tabularx}
\end{table*}

\begin{figure*}[t]
    \centering

    \begin{subfigure}{0.32\textwidth}
        \centering
        \includegraphics[width=\linewidth]{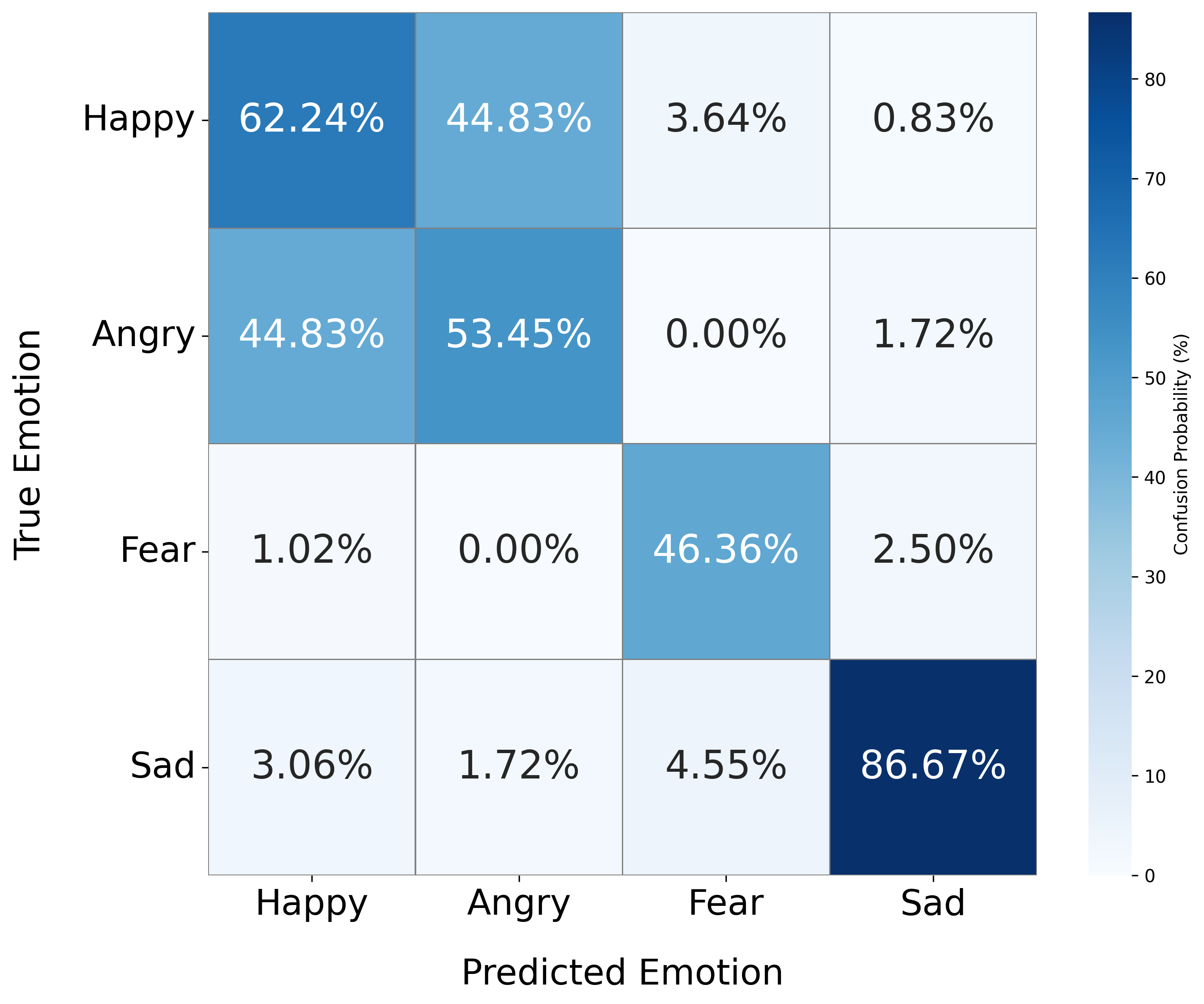}
        \subcaption{Only verbal}
        \label{fig:sub2}
    \end{subfigure}
    \hfill 
    \begin{subfigure}{0.32\textwidth}
        \centering
        \includegraphics[width=\linewidth]{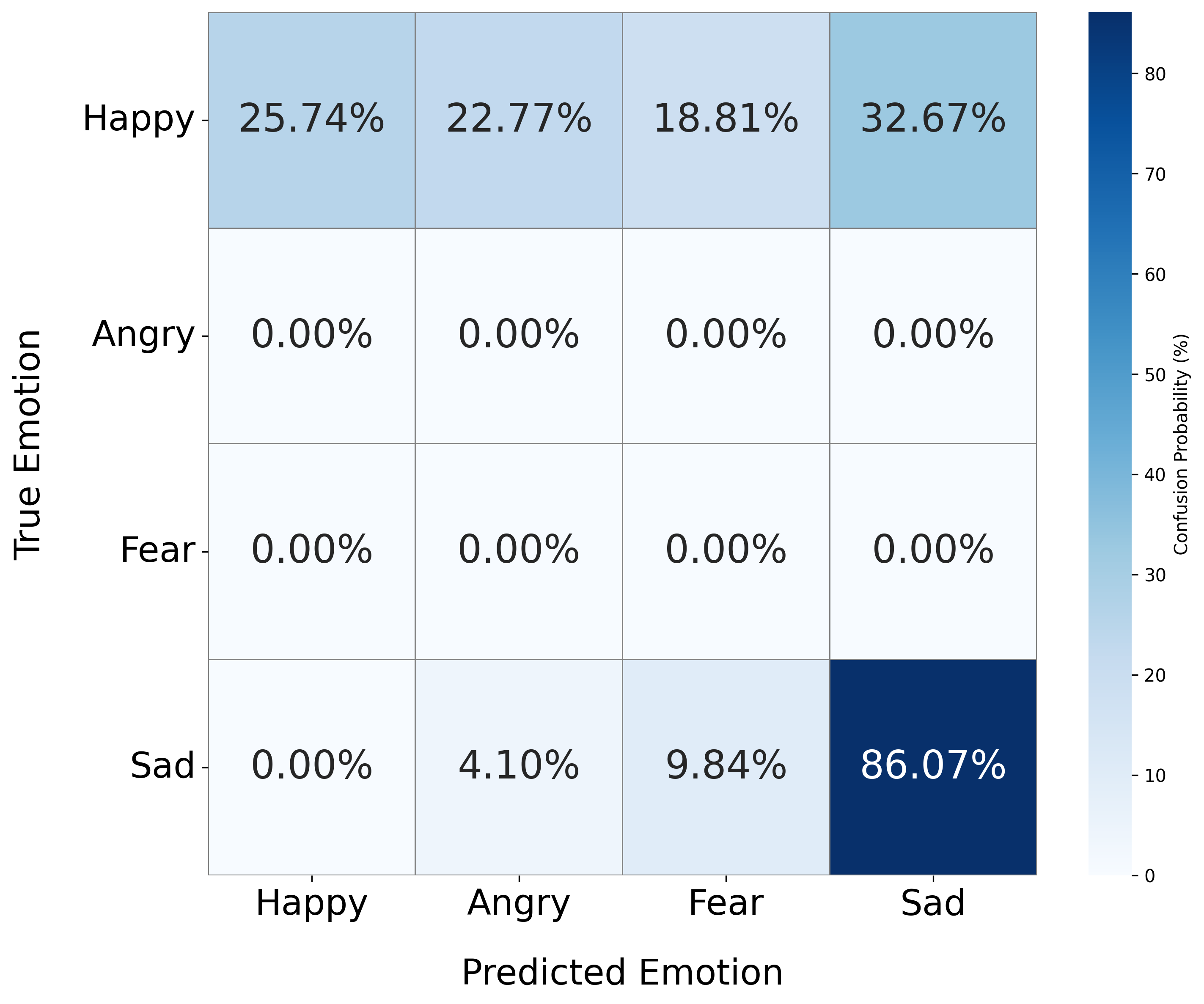}
        \subcaption{Coarse-grained non-verbal}
        \label{fig:sub3}
    \end{subfigure}
    \hfill 
    \begin{subfigure}{0.32\textwidth}
        \centering
        \includegraphics[width=\linewidth]{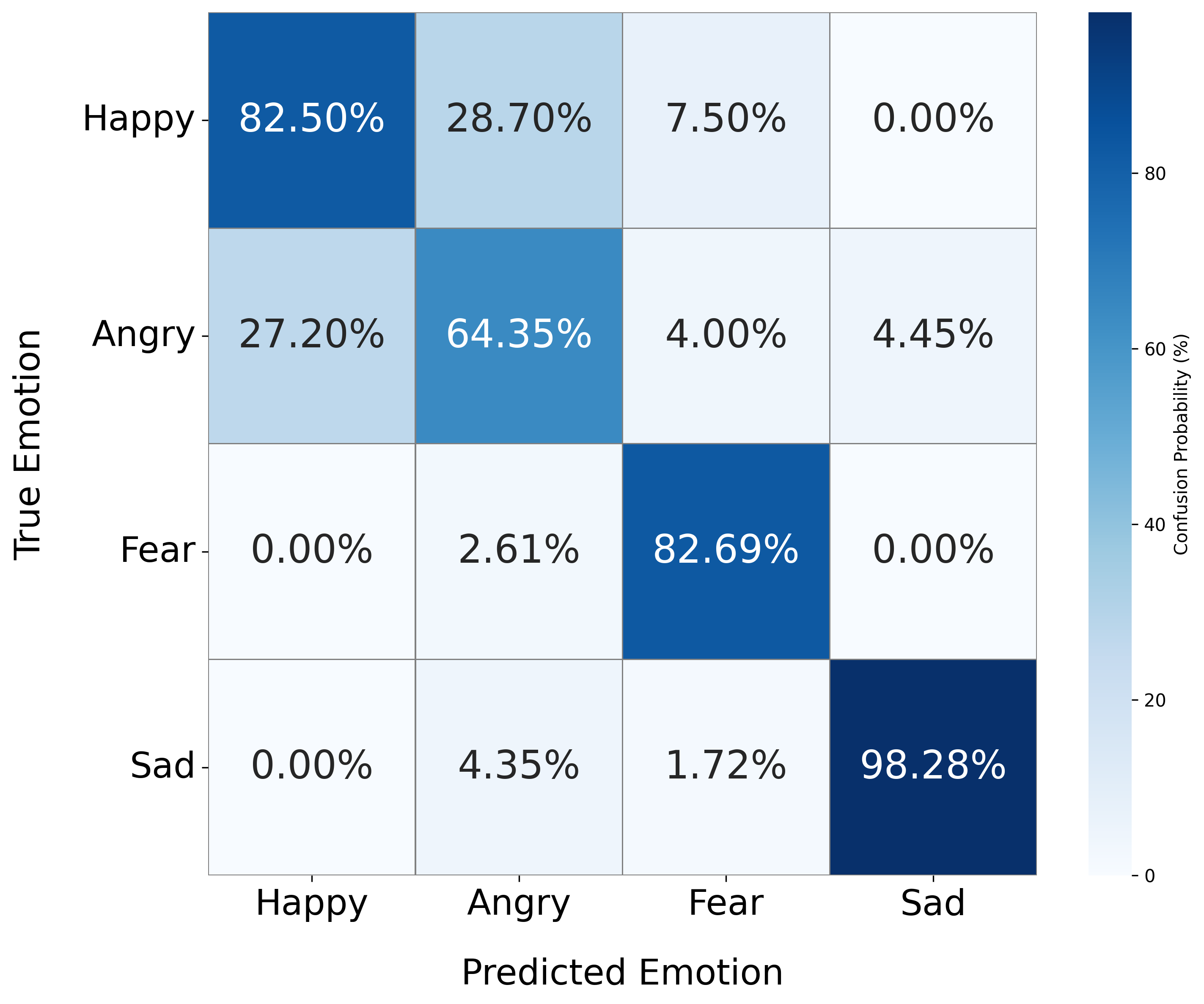}
        \subcaption{Fine-grained non-verbal}
        \label{fig:sub1}
    \end{subfigure}
    
    \caption{Emotion recognition confusion matrices for (a) Only verbal, (b) Verbal + coarse-grained non-verbal, and (c) Verbal + fine-grained non-verbal.}
    \label{fig:three_figures_comparison}
\end{figure*}

\subsubsection{Emotion-Specific Analysis}
\label{sec:specific}

To provide a more detailed breakdown of each emotion, we conducted an analysis across individual emotion categories. Figure\ref{fig:three_figures_comparison} and table\ref{tab:detailed_results_final_vline} present the accuracy, nMOS, and eMOS of each emotion.

\noindent\textbf{Happy}: Our proposed design achieved the highest accuracy at 82.5\%, significantly outperforming the other two. The particularly low score of the `Coarse-grained NV' suggests that its happy-related non-verbal cues (laugh) were perceived as ambiguous, its laughter samples were often quiet and subdued, leading many listeners to mistakenly perceive the emotion as 'sad' rather than 'happy'.

\noindent\textbf{Sad}: All three designs performed well for sad emotions, with our proposed design achieving the highest accuracy of 98.3\%. This indicates that sad emotions are relatively easier to recognize from both verbal and non-verbal cues. However, the subtle differences in scores suggest that our non-verbal cues, which include a variety of crying sounds, still provide a slight edge in conveying sadness compared to the other methods.

\noindent\textbf{Angry}: Our method achieved 64.3\% accuracy, a modest improvement compared to other emotions. We hypothesize that this is because we lacked a direct non-verbal cue that is uniquely associated with anger. While sad emotions are immediately recognizable through a crying NV, our primary NV for anger was 'yelling'. Yelling, however, is a more general high-arousal expression and does not instantly lead listeners to a specific 'angry' conclusion. This may be why its accuracy did not see a dramatic increase and why its E-MOS score was 4.06, which is lower than the 4.2+ scores of other emotions.

\noindent\textbf{Fear}: Our method reached a remarkable 82.7\% accuracy, marking an approximately 36\% improvement over the 'Only Verbal'. This highlights the crucial role of non-verbal cues (screaming) in expressing fear.

\subsubsection{Preference Results}
\begin{figure}[h!] 
    \centering

    \begin{subfigure}{\linewidth}
        \centering
        \includegraphics[width=\linewidth]{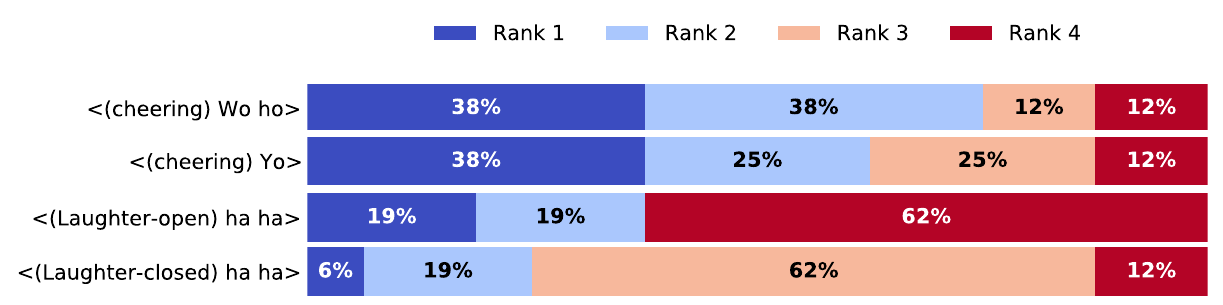}
        \subcaption{The results for the Happy emotion.}
        \label{fig:happy_results}
    \end{subfigure}
    
    \vspace{5mm} 
    
    \begin{subfigure}{\linewidth}
        \centering
        \includegraphics[width=\linewidth]{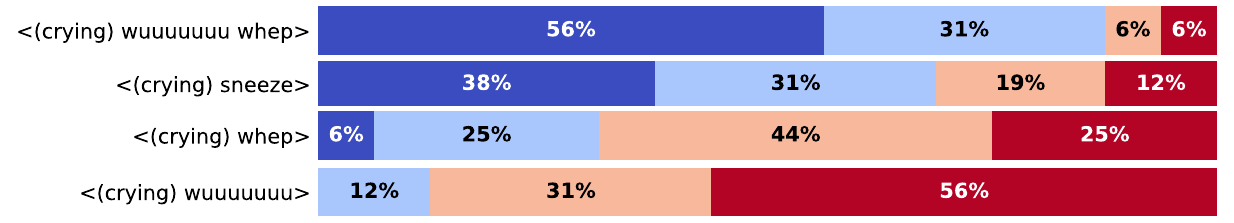}
        \subcaption{The results for the Sad emotion.}
        \label{fig:sad_results}
    \end{subfigure}
    
    \caption{Preference evaluation results for different non-verbal expressions.}
    \label{fig:mos_combined}
\end{figure}

The results of our preference evaluation are visualized in Figures \ref{fig:happy_results} and \ref{fig:sad_results}.

For the Happy emotion, cheering sounds were distinctly preferred over laughter. Both \texttt{<(cheering) Wo ho>} and \texttt{<(cheering) Yo>} were ranked as the top preference by a large portion of participants, indicating that cheering is perceived as a more effective way to convey joy and excitement. In contrast, laughter expressions, especially \texttt{<(Laughter-closed) ha ha>}, were consistently ranked lower, with 62\% of participants ranking them third.

For the Sad emotion, a complex, multi-part expression was significantly preferred. The expression \texttt{<(crying) wuuuuuuu whep>} (combining a prolonged cry with a sob) was ranked first by 56\% of participants. This suggests that a nuanced non-verbal expression with multiple elements is perceived as the most authentic representation of sadness. Simple, single-sound expressions like \texttt{<(crying) whep>} and \texttt{<(crying) wuuuuuuu>} received lower preference ratings, with the latter being ranked fourth by 56\% of participants.

These results provide valuable guidance for our fine-grained non-verbal design, validating the superiority of specific, structured non-verbal vocalizations in conveying particular emotions.


\section{Conclusion}
In this paper, we constructed a Fine-Grained Non-Verbal Expression Data and built a Fine-Grained Non-Verbal emotional TTS benchmark to enable the use of fine-grained non-verbal cues. Our comprehensive subjective evaluation yielded three key findings. 
First, while the inclusion of NVs resulted in a minor trade-off in perceived speech naturalness compared to `verbal-only’, this was outweighed by significant gains in other metrics. Second, our ``Fine-Grained Non-Verbal" design achieved superior performance in both Expressiveness and Emotional Recognition Accuracy, with an average E-MOS score of 4.20 and average accuracy of 82.0\%. This demonstrated that our designed ``Fine-Grained Non-Verbal" is more effective than the coarse-grained NVs found in other corpora. Third, our emotion-specific analysis confirmed that our non-verbal cues were particularly effective in conveying emotions. They were especially potent for high-arousal emotions like happy (82.5\% accuracy) and fear (82.7\% accuracy). Furthermore, the accuracy for sad emotions was exceptionally high, achieving 98.3\%. These results indicate that our non-verbal cues provided a distinct emotional manifestation and burst that was significantly more effective than what could be achieved with verbal-only prosody.
Finally, our preference test provided valuable insights into listener expectations, revealing a strong preference for cheering over laughter for happy emotions and for multi-part crying expressions for sad emotions. These findings validate our approach and highlight the importance of not just including NVs, but also designing them with granularity and user preference in mind to achieve truly expressive and human-like emotional synthesis.

\vspace{3mm} 
\noindent\textbf{Acknowledgement} 
Part of this work was supported by JSPS KAKENHI Grant Numbers JP21H05054, JP23K21681, JP24K0296, and JP25H01139, as well as JST NEXUS (JPMJNX25C1).


\bibliographystyle{IEEEbib}
\bibliography{strings,refs}

\end{document}